\documentclass[11pt,a4paper]{article}
\usepackage{amsmath,amssymb}
\usepackage{epsfig,graphicx}
\usepackage{epic,eepic}
\usepackage{subfigure}
\usepackage{rotating}
\usepackage{bm}
\usepackage{axodraw4j}
\usepackage{color}
\usepackage{hyperref}
\usepackage{cite}

\newlength{\wth}
\def\Tr{{\rm Tr}}

\def\Dbarslash{\,\,{\raise.15ex\hbox{/}\mkern-12mu {\bar\D}}}
\def\Dslash{\,\,{\raise.15ex\hbox{/}\mkern-12mu \D}}
\def\delslash{\,\,{\raise.15ex\hbox{/}\mkern-9mu \partial}}
\def\delbarslash{\,\,{\raise.15ex\hbox{/}\mkern-9mu {\bar\partial}}}

 \def\L{\Lambda}

\def\D{{\cal D}}
\def\Dbarslash{\,\,{\raise.15ex\hbox{/}\mkern-12mu {\bar\D}}}
\def\delslash{\,\,{\raise.15ex\hbox{/}\mkern-9mu \partial}}
\def\Dslash{\,\,{\raise.15ex\hbox{/}\mkern-12mu \D}}

\def\={\, =\, }
\def\+{\, +\, }
\def\-{\, -\, }

\def\slashchar#1{\setbox0=\hbox{$#1$}           
\dimen0=\wd0                                 
\setbox1=\hbox{/} \dimen1=\wd1               
\ifdim\dimen0>\dimen1                        
   \rlap{\hbox to \dimen0{\hfil/\hfil}}      
   #1                                        
\else                                        
   \rlap{\hbox to \dimen1{\hfil$#1$\hfil}}   
   /                                         
\fi}

\newcommand{\be}{\begin{equation}}
\newcommand{\ee}{\end{equation}}
\def\bea{\begin{eqnarray}}
\def\eea{\end{eqnarray}}

\definecolor{saabeer}{rgb}{0,1,0}

\definecolor{durbeer}{rgb}{1,0,0}

\definecolor{durbeer2}{rgb}{0,0,0.7}

\begin{document}
\date{\mbox{ }}
\title{{\normalsize  IPPP/11/08; DCPT/11/16\hfill\mbox{}\hfill\mbox{}}\\
\vspace{2.5cm} \LARGE{\textbf{RG Invariants, Unification and
the Role  of the Messenger Scale in General Gauge Mediation}}}
\author{
Joerg Jaeckel, Valentin V. Khoze and Chris Wymant\\[4ex]
\small{\em Institute for Particle Physics Phenomenology, Department of Physics,}\\
\small{\em Durham University, Durham DH1 3LE, United Kingdom}\\[2ex]
}  
\date{}
\maketitle

\vspace{3ex}

\begin{abstract}
\noindent
In General Gauge Mediation (GGM) {\it all} MSSM soft sfermion masses at a high scale $M_{\rm mess}$ can be parameterised  by three a priori independent scales $\Lambda_{S;\, 1,2,3}(M_{\rm mess})$. (Similarly the gaugino masses are given by $\Lambda_{G;\, 1,2,3}(M_{\rm mess})$.)
For the first two generations this parameterisation in terms of a set of running $\Lambda_{S;\, 1,2,3}(\mu)$ -- conveniently obtained from appropriate RG invariants -- continues to hold all the way down to the electroweak scale. This is not the case for the third generation because of the large Yukawa couplings.
Together these two observations imply that the messenger scale is an additional parameter of GGM models. \\
In models where all messengers are in complete GUT multiplets (without significant mass splittings), all $\Lambda_{S, r}$ are equal at $M_{\rm mess}$.
Starting from the observable mass spectrum at the electroweak scale we present a strategy to determine if this unification occurs and at
which scale. This approach uses data accessible at colliders to gain insight into high scale unification physics beyond the unification of gauge couplings.\\
\end{abstract}

\newpage


\section{Introduction}
In the gauge mediation framework Standard Model gauge interactions transmit supersymmetry (SUSY) breaking effects from
the hidden sector to the Standard Model superpartners.
In most realisations this generates SUSY breaking soft terms by integrating out messenger fields at the messenger scale $M_{\rm mess}$.
This scale is typically much higher than the electroweak scale where physics is probed at colliders.

For early implementations of the gauge mediation idea we refer the reader to~\cite{GRat}. More recently, it was discovered in the context of explicit models~\cite{Izawa:1997gs,Kitano:2006xg,Csaki:2006wi,Abel:2007jx,Haba:2007rj,Abel:2007nr,Abel:2008gv,Komargodski:2009jf,Abel:2009ze,Cheung:2007es} that the parameter space
is much richer than originally envisaged.
This is conveniently embraced in the General Gauge Mediation framework introduced in~\cite{GGM,Buican:2008ws}.
The phenomenology of these models has been studied in~\cite{ADJKpGGM,ADJK7,Carpenter:2008he,Rajaraman:2009ga,Kobayashi:2009rn,Meade:2010ji,Ruderman:2010kj,Thalapillil:2010ek}.

For gaugino and sfermion masses of MSSM this leads to the following general structure
\bea
\label{gauginosoft}
M_{r}(M) &=&\, k_r \,\frac{\alpha_r}{4\pi}\,\,\Lambda_{G,r} \, , \\
\label{scalarsoft}
m_{\tilde{f}}^2 (M) &=&\, 2 \sum_{r=1}^3 C_2(f,r) k_r \,\frac{\alpha_r^2}{(4\pi)^2}\,\, \Lambda_{S,r}^2\, ,
\eea
where $k_r = (5/3,1,1)$, $\alpha_r (M)$ are the gauge coupling constants\footnote{For $g_{1}$ we use the Standard Model rather than GUT normalisation.} 
and $C_2(f,r)$ are the quadratic Casimir operators of the representation of $f$ under the $r^{\rm th}$ gauge group,
$C_{2}(3)=4/3$, $C_{2}(2)=3/4$, $C_{2}(1)=Y^2$ for sfermions in the fundamental representation ($C_2=0$ for singlets of a gauge group).
Ordinary gauge mediation scenarios occupy the restricted
parameter space $\L_G\simeq\L_S$.

These equations parameterise the five sfermion massses for each generation in terms of only three parameters $\Lambda_{S,r}$.
This requires two linear combinations of sfermion mass squareds to vanish, which amounts to two mass sum rules. These can be chosen~\cite{GGM}
as $\Tr[(B-L)m^2]=0$ and $\Tr(Ym^2)=0$. In a recent work~\cite{Jaeckel:2011ma} we demonstrated that these sum rules hold to excellent accuracy
for the first two generations all the way down to the electroweak scale\footnote{Assuming universal Higgs masses, $\delta_{u}=\delta_{d}=0$ 
in the language of Eq.~\eqref{higgssoft}.}. For the third generation there are significant violations of the hypercharge sum rule. The amount by which this sum rule is violated at the electroweak scale depends on the messenger scale. This already points to the fact that the messenger scale $M_{\rm mess}$ is an additional parameter that has to be included in the description of GGM model space. Later in this paper we will make this notion more concrete.

Equations~\eqref{gauginosoft} and \eqref{scalarsoft} are derived and expected to hold at a certain high scale $M$. In common models
of gauge mediation with explicit messenger fields it is the messenger scale $M_{\rm mess}$.
The standard top-down approach is to use Eqs.~\eqref{gauginosoft} and \eqref{scalarsoft} along with other soft terms - if non-vanishing -
as input parameters at the high scale which are subsequently RG evolved down to the electroweak scale using, for example,
{\texttt{SoftSUSY}}~\cite{Allanach:2001kg}. In the context
of pure GGM\footnote{Pure GGM is the strict definition of gauge mediation where all soft parameters in the MSSM vanish in the limit where all Standard Model gauge couplings are zero.} this approach was followed and phenomenologically studied in~\cite{ADJKpGGM,ADJK7}.

In this paper we also use a bottom up approach which starts with the low energy spectrum, with the aim of uncovering essential features of the underlying theory at high energies. Specifically we will use the set of one-loop RG invariants assembled in~\cite{Carena:2010gr,Carena:2010wv},
as a tool to investigate the theory at energy scales $\sim M_{\rm mess}$, well above the scale which can be directly probed at colliders.

In the following Sect.~\ref{rgi} we briefly review the one-loop RG invariants of the MSSM in the context of GGM.
Then in Sect.~\ref{parameter} we use these tools to elucidate the role of $M_{\rm mess}$ as an independent parameter of GGM models.
In Sect.~\ref{sec:unification} we will present a strategy to discover whether the underlying theory exhibits a unification of the $\Lambda$ parameters for the three gauge groups. If so, this indicates a unification principle which goes beyond gauge coupling unification. At the same time it gives the value of the messenger scale where the three $\Lambda_{S}$ parameters coincide. This scale has the meaning of a mass of messenger fields transforming in a complete and un-split GUT multiplets.
As this $\Lambda$-unification assumes the specific way the $\Lambda$s appear in Eqs.~\eqref{gauginosoft} and \eqref{scalarsoft}, if observed it would also point to gauge mediation rather than other types of mediation such as via gravity.

\section{1-loop RG invariants}\label{rgi}

In this section we briefly review the renormalization group invariants (RGIs) as presented in~\cite{Carena:2010gr}.

In the MSSM there are a number of quantities which do not change under 1-loop RG evolution.
In other words
\begin{equation}
\label{rgidef}
\frac{d}{dt}({\rm{RGI}})=0+{\mathcal{O}}({\text{2-loop}}),
\end{equation}
with $t = \log (\mu / M_{\rm mess})$.
The 1-loop RGIs of~\cite{Carena:2010gr} are listed in the first column of Tab.~\ref{rginvariants},
their definition in terms of MSSM parameters in the second column.

In the third column of Tab.~\ref{rginvariants} we give the values for the RGIs in GGM using the parametrisation of the soft masses
given in Eqs.~\eqref{gauginosoft} and \eqref{scalarsoft}.
If we employed the strict definition of pure GGM, i.e. all soft masses vanish in the limit that the Standard Model gauge couplings are zero, this would be a complete parameterisation of the model. In this setup $B_{\mu}\approx 0$ at the high scale and $\tan\beta$ is not a parameter but a prediction~\cite{ADJKpGGM}.
However, in a less strict definition of GGM one often allows for additional couplings of the Higgs fields to the SUSY breaking sector. 
Here, a non-vanishing $B_{\mu}$ is already generated at the high scale, which in turn gives a
value of (and can be traded for) $\tan\beta$ at the low scale. For practical computations it is, however, often convenient to use $\tan\beta$ directly as an input parameter (instead of $B_{\mu}$).
The introduction of new couplings to the Higgs field also generates
contributions, $\delta_{u}$ and $\delta_{d}$, to the Higgs soft mass parameters,
\begin{equation}
\label{higgssoft}
m^{2}_{H_{u}}=m^{2}_{\tilde{L}}+\delta_{u}, \quad\quad m^{2}_{H_{d}}=m^{2}_{\tilde{L}}+\delta_{d}.
\end{equation}
This less strict definition of GGM thus has three additional parameters -- $\tan\beta$, $\delta_{u}$ and $\delta_{d}$ --
compared to the six $\Lambda$s in pure GGM.

A number of the RGIs collected in Tab.~\ref{rginvariants} contain first generation soft masses but there are none containing the corresponding second generation terms. To the level of approximation employed here and in~\cite{Carena:2010gr} one does not distinguish between the first and second generation sfermions (both having small Yukawa couplings). Therefore, replacing the first with the second generation in these RGIs would just give a set of identical RGIs and yields no additional information.

\begin{figure}[t]
\centering
\begin{picture}(370,200)
\includegraphics[width=0.8\linewidth]{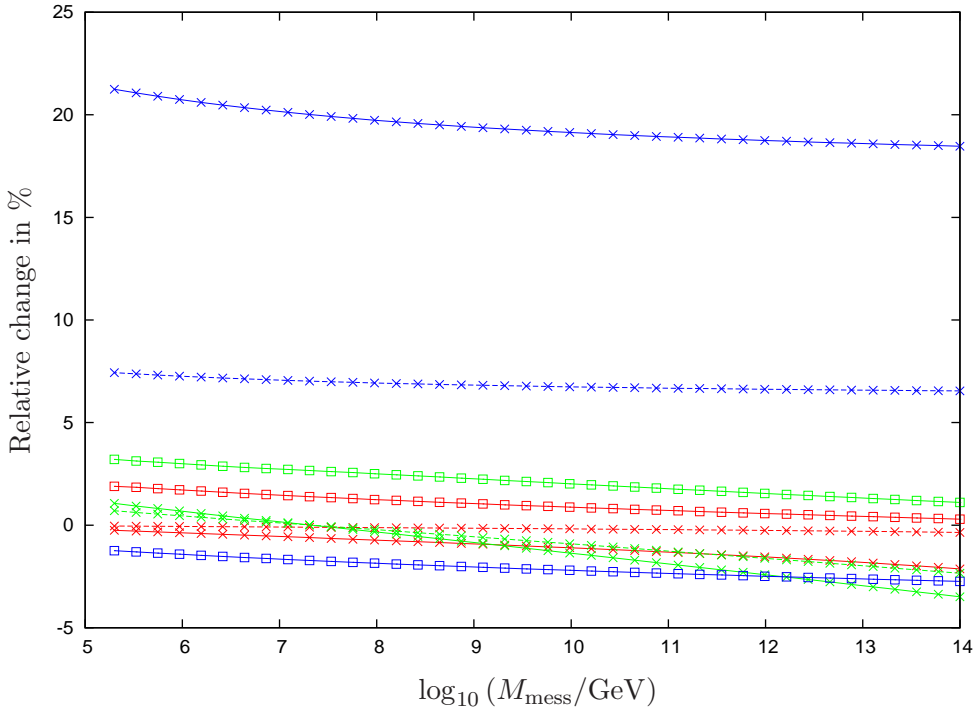}
\Text(-170,-10)[c]{\scalebox{1}[1]{$\log_{10} \left(M_{\rm mess}/{\rm GeV}\right)$}}
\rText(-365,130)[c][l]{\scalebox{1}[1]{Relative change in \%}}
\end{picture}
\vspace{0.4cm}
\caption{Difference between the value of the RGI at the messenger scale $M_{\rm mess}$ and at the low scale, relative to the value
of the RGI itself at $M_{\rm mess}$ (solid lines) or relative to the largest single mass squared term appearing in the definition on the 
RGI (dashed). Boxes denote the $I_{B_{r}}$, crosses the $I_{M_{r}}$. Red, green and blue correspond to the gauge groups $r=1$, 2 and 3.
We have varied the messenger scale keeping the values all RGIs fixed at $M_{\rm mess}$. Specifically we have chosen values such that
all six $\Lambda$s are equal, $\Lambda_{S,r}=\Lambda_{G_r}=10^5\, {\rm GeV}$, at the
messenger scale $M_{\rm mess}=10^{14}\, {\rm GeV}$. When the messenger scale is changed the $\Lambda_{S}$ values at $M_{\rm mess}$ change as in Fig.~\ref{unification} and as discussed in more detail in Sect.~\ref{parameter}. $\tan\beta=45$.}
\label{fig:rgicheck}
\end{figure}

Strictly speaking these RGIs are exact only at one loop. We have compared the values of the RGIs at the high and the low scales numerically,
taking into account 2-loop effects using {\texttt{SoftSUSY 3.1.6}}~\cite{Allanach:2001kg}.
Fig.~\ref{fig:rgicheck} shows the difference between the value of the RGI at the messenger scale and at the low electroweak scale
for a range of messenger masses, keeping the value of the RGI at $M_{\rm mess}$ fixed.
We can see that the RGIs remain constant to a reasonably good accuracy ($\lesssim 10\%$ of the value of the RGI over the explored parameter space). 
The somewhat larger relative change in $I_{M_{3}}$ is due to a partial cancellation between 
$\Lambda_{G}$ and $\Lambda_{S}$ at the high scale, as one can see from Tab.~\ref{rginvariants}. Importantly, the absolute change remains reasonably small, as can be seen when comparing the change in the RGIs to the size of the individual soft masses contributing 
(dashed lines in Fig.~\ref{fig:rgicheck}). 

\begin{table}[!hp]
\centering
\begin{tabular}{| c | c | c |}
\hline
\hline
&&\\
RGI &  Definition in terms Soft Masses & GGM value  \\ [0.8ex]
 & &  \\
\hline
\hline
&&\\
$D_{B_{13}}$&$2(m_{\tilde{Q}_1}^2-m_{\tilde{Q}_3}^2)-m_{\tilde{u}_1}^2+m_{\tilde{u}_3}^2-m_{\tilde{d}_1}^2+m_{\tilde{d}_3}^2$
& 0 \\ [3ex]
\hline
&&\\
$D_{L_{13}}$&$2(m_{\tilde{L}_1}^2-m_{\tilde{L}_3}^2)-m_{\tilde{e}_1}^2+m_{\tilde{e}_3}^2$& 0\\ [3ex]
\hline
&&\\
$D_{\chi_1}$&$3(3m_{\tilde{d}_1}^2-2(m_{\tilde{Q}_1}^2-m_{\tilde{L}_1}^2)-m_{\tilde{u}_1}^2)-m_{\tilde{e}_1}^2$& 0\\ [3ex]
\hline&&\\
$D_{Y_{13H}}$&$\begin{array}{c} m^2_{\tilde{Q}_1}-2m^2_{\tilde{u}_1}+m^2_{\tilde{d}_1}-m^2_{\tilde{L}_1}+m^2_{\tilde{e}_1}\\-\frac{10}{13}\left(m^2_{\tilde{Q}_3}-2m^2_{\tilde{u}_3}+m^2_{\tilde{d}_3}-m^2_{\tilde{L}_3}+m^2_{\tilde{e}_3}+m^2_{H_u}-m^2_{H_d}\right)\end{array}$
&$-\frac{10}{13}(\delta_u-\delta_d)$\\ [6ex]
\hline
&&\\
$D_{Z}$&$3(m_{\tilde{d}_3}^2-m_{\tilde{d}_1}^2)+2(m_{\tilde{L}_3}^2-m_{H_d}^2)$ & $-2\delta_{d}$\\ [3ex]
\hline
&&\\
$I_{Y\alpha}$&$\left(m^2_{H_u}-m^2_{H_d}+\sum_{gen}(m^2_{\tilde{Q}}-2m^2_{\tilde{u}}+m^2_{\tilde{d}}-m^2_{\tilde{L}}+m^2_{\tilde{e}})\right)/g_1^2$
&$\left(\delta_u-\delta_d\right)/g_1^2$\\ [3ex]
\hline
&&\\
$I_{B_r}$&$M_r/g_r^2$&$\frac{k_{r}\Lambda_{G,r}}{16\pi^2}$ \\ [3ex]
\hline
&&\\
$I_{M_1}$&$M_1^2-\frac{33}{8}(m_{\tilde{d}_1}^2-m_{\tilde{u}_1}^2-m_{\tilde{e}_1}^2)$&
$\frac{25}{9}\frac{g^{4}_{1}(M_{\rm mess})}{(16\pi^2)^{2}}\left(\Lambda_{G,1}^{2}+\frac{33}{5}\Lambda^{2}_{S,1}\right)$\\ [3ex]
\hline
&&\\
$I_{M_2}$&$M_2^2+\frac{1}{24}\left(9(m_{\tilde{d}_1}^2-m_{\tilde{u}_1}^2)+16m_{\tilde{L}_1}^2-m_{\tilde{e}_1}^2\right)$
&$\frac{g^{4}_{2}(M_{\rm mess})}{(16\pi^2)^{2}}\left(\Lambda_{G,2}^{2}+\Lambda^{2}_{S,2}\right)$\\ [3ex]
\hline
&&\\
$I_{M_3}$&$M_3^2-\frac{3}{16}(5m_{\tilde{d}_1}^2+m_{\tilde{u}_1}^2-m_{\tilde{e}_1}^2)$&
$\frac{g^{4}_{3}(M_{\rm mess})}{(16\pi^2)^{2}}\left(\Lambda_{G,3}^{2}-3\Lambda^{2}_{S,3}\right)$\\ [3ex]
\hline
&&\\
$I_{g_2}$&$ \frac{3}{5g_1^2}-\frac{33}{5g_2^2}$&$\approx -10.9$\\ [3ex]
\hline
&&\\
$I_{g_3}$&$ \frac{3}{5g_1^2}+\frac{33}{15g_3^2}$&$\approx 6.2$\\ [3ex]
\hline
\end{tabular}
\caption{1-loop RG invariant quantities (taken and adapted from~\cite{Carena:2010gr}). The third column gives their value in GGM models using the connection to soft masses given in Eqs.~\eqref{gauginosoft}, \eqref{scalarsoft} and \eqref{higgssoft}.}
\label{rginvariants}
\end{table}

We also note that of the RGIs given in Tab.~\ref{rginvariants}, $D_{\chi_{1}}$ and $I_{M_{r}}$ are not strictly speaking RGIs in the sense of Eq.~\eqref{rgidef}: they are only RGIs if the first (and second) generation Yukawa couplings are neglected. This explains why there
are no corresponding third generation RGIs.
Indeed, using the RG equations for the soft masses~\cite{Martin:1997ns} we can write down the evolution for the analogs of the $I_{M_r}$, 
which we call $I_{M_{r},3}$, defined with third generation in place of first generation sfermion masses,
\begin{equation} 
\label{RGvariants}
\begin{split}
16\pi^2 \frac{d}{dt} I_{M_1,3} & = \frac{33}{4}\left( X_t - X_b + X_{\tau} \right) \\
16\pi^2 \frac{d}{dt} I_{M_2,3} & = \frac{1}{12}\left( -9 X_t + 9 X_b + 7 X_{\tau} \right) \\
16\pi^2 \frac{d}{dt} I_{M_3,3} & = \frac{3}{8}\left( -X_t - 5 X_b + X_{\tau} \right)
\end{split}
\end{equation}
\begin{equation*}
\begin{split}
\text{with} \qquad X_t & \equiv 2 |y_t|^2      (m^2_{H_u} + m_{\tilde{Q}_{3}}^{2} + m_{\tilde{u}_{3}}^{2}) + 2 |a_t|^2 \\
X_b & \equiv 2 |y_b|^2 (m^2_{H_d} + m_{\tilde{Q}_{3}}^{2} + m_{\tilde{d}_{3}}^{2}) + 2 |a_b|^2 \\
X_{\tau} & \equiv 2 |y_{\tau}|^2 (m^2_{H_d} + m_{\tilde{L}_{3}}^{2} + m_{\tilde{e}_{3}}^{2}) + 2 |a_{\tau}|^2.
\end{split}
\end{equation*}
The right hand sides of these equations are non-vanishing (and not small).
In Sect.~\ref{parameter} we will use the resulting running of these ``RGIs'' to argue that the messenger scale cannot be ignored and plays the role
of a true parameter in GGM.

Finally let us also comment on the relation between RGIs and the sum rules given in~\cite{GGM} and investigated in some
detail in our earlier paper~\cite{Jaeckel:2011ma}.
In GGM the following sum rules for soft masses hold at the messenger scale for each individual generation $i$ (as can be easily checked
by using the parameterisation \eqref{gauginosoft} and \eqref{scalarsoft}),
\bea
\label{sumrules}
S_{Y,i}=\Tr(Ym^2_{i})=m_{\tilde{Q}^2_{i}} -2m_{\tilde{u}^2_{i}}+m_{\tilde{d}^2_{i}}-m_{\tilde{L}^2_{i}}+m_{\tilde{e}^2_{i}} &=& 0 \, ,  \\\nonumber
S_{B-L,i}=\Tr((B-L)m^2_{i})=2m_{\tilde{Q}^2_{i}} -m_{\tilde{u}^2_{i}}-m_{\tilde{d}^2_{i}}-2m_{\tilde{L}^2_{i}}+m_{\tilde{e}^2_{i}} &=& 0 .
\eea

If we neglect $\delta_{u}$ and $\delta_{d}$, the first six RGIs in Tab.~\ref{rginvariants} evaluate to zero and can therefore be called a sum
rule. One can see that these sum rules are related to the standard GGM sum rules Eq.~\eqref{sumrules}.
Indeed we have,
\begin{eqnarray}
\label{bminusl}
&&D_{B_{13}}-D_{L_{13}}= S_{B-L,1}-S_{B-L,3},\\
\label{hypercharge}
&&g^2_{1}I_{Y\alpha}=m^{2}_{H_{u}}-m^{2}_{H_{d}}+S_{Y,1}+S_{Y,2}+S_{Y,3}.
\end{eqnarray}
The RGIs on the left hand side are invariants even when Yukawa couplings are included. As we have checked in~\cite{Jaeckel:2011ma}
$S_{B-L,1}$ itself is a good sum rule as it holds also at the electroweak scale. Using now that \eqref{bminusl} is an RGI we see that
$S_{B-L,3}=0$ also must hold at the low scale to good accuracy\footnote{For the third generation there is a small subtlety. The sum rules and the RGIs are written in terms of soft mass parameters. The sfermion masses of the third generation are also significantly affected by electroweak symmetry breaking. In order to evaluate the sum rules one has to extract the soft mass parameters from the particle masses and the mixing angle between left and right handed sfermions~\cite{Jaeckel:2011ma}.} . This was explicitly confirmed in~\cite{Jaeckel:2011ma}.
Similarly, for the first two generations $S_{Y,i}\approx0$ at all scales. Using the RGI \eqref{hypercharge} we can
see that only a combination of $S_{Y,3}$ and Higgs mass parameters can form a good sum rule at the low scale.

\section{$M_{\rm mess}$ is a parameter of GGM}\label{parameter}
Having confirmed that the RGIs are indeed constant during the RG evolution we can now try to use them as a convenient parameterisation
of the GGM parameter space. Indeed there are six non-vanishing
RGIs, $I_{B_{r}}$ and $I_{M_{r}}$. For a given messenger scale, fixing their values uniquely determine all six $\Lambda$s in our GGM parameterisation
\eqref{gauginosoft}, \eqref{scalarsoft}.

One can now wonder whether GGM models with the same values for these six RGIs but different values of the messenger mass give
the same physical spectra. As we will see this is not the case. Consequently $M_{\rm mess}$ is a true parameter of GGM
which has to be taken into account in addition to the six $\Lambda$s.

\begin{figure}[!hp]
\centering
\subfigure{
\begin{picture}(350,260)
\includegraphics[width=0.8\linewidth]{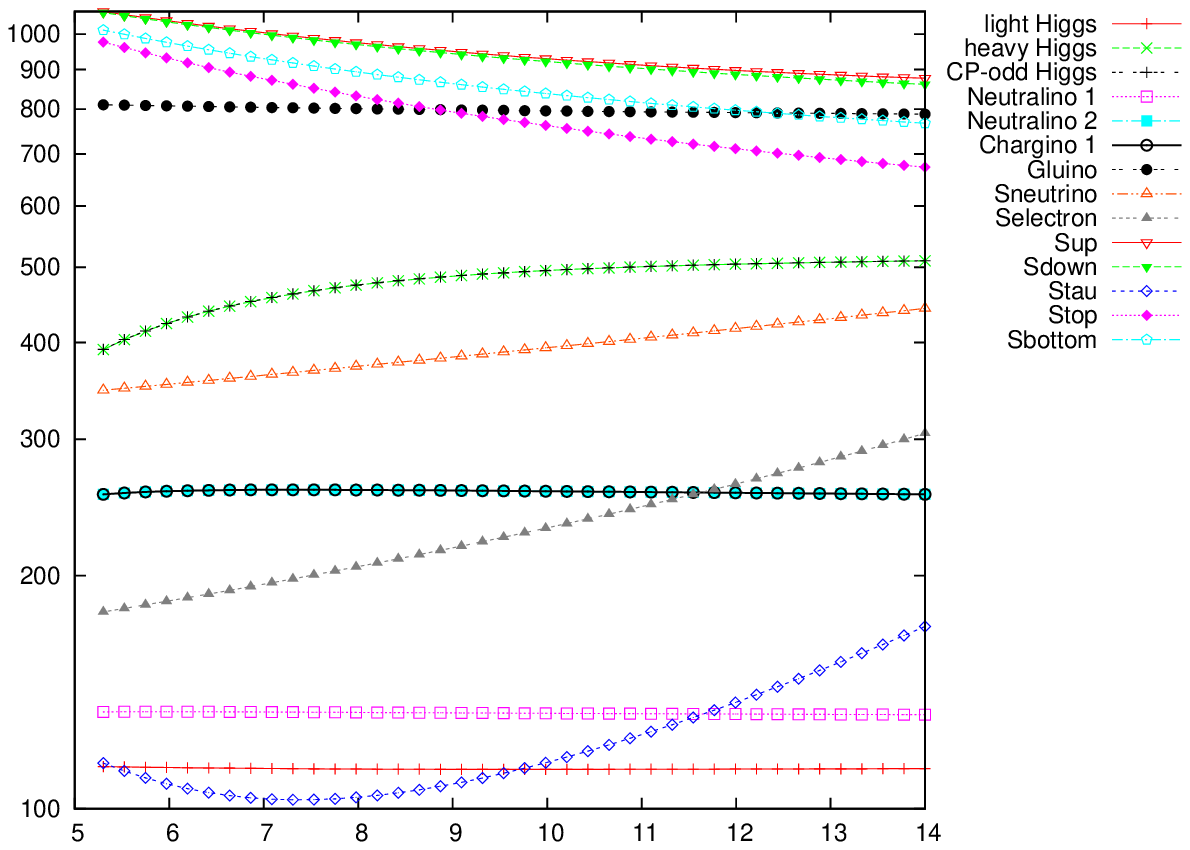}
\rText(-365,150)[c][l]{\scalebox{1}[1]{Mass in GeV}}
\Text(-200,-10)[c]{\scalebox{1}[1]{$\log_{10} \left(M_{\rm mess}/{\rm GeV}\right)$}}
\end{picture}
}
\\[0.7cm]
\subfigure{
\begin{picture}(350,260)
\includegraphics[width=0.8\linewidth]{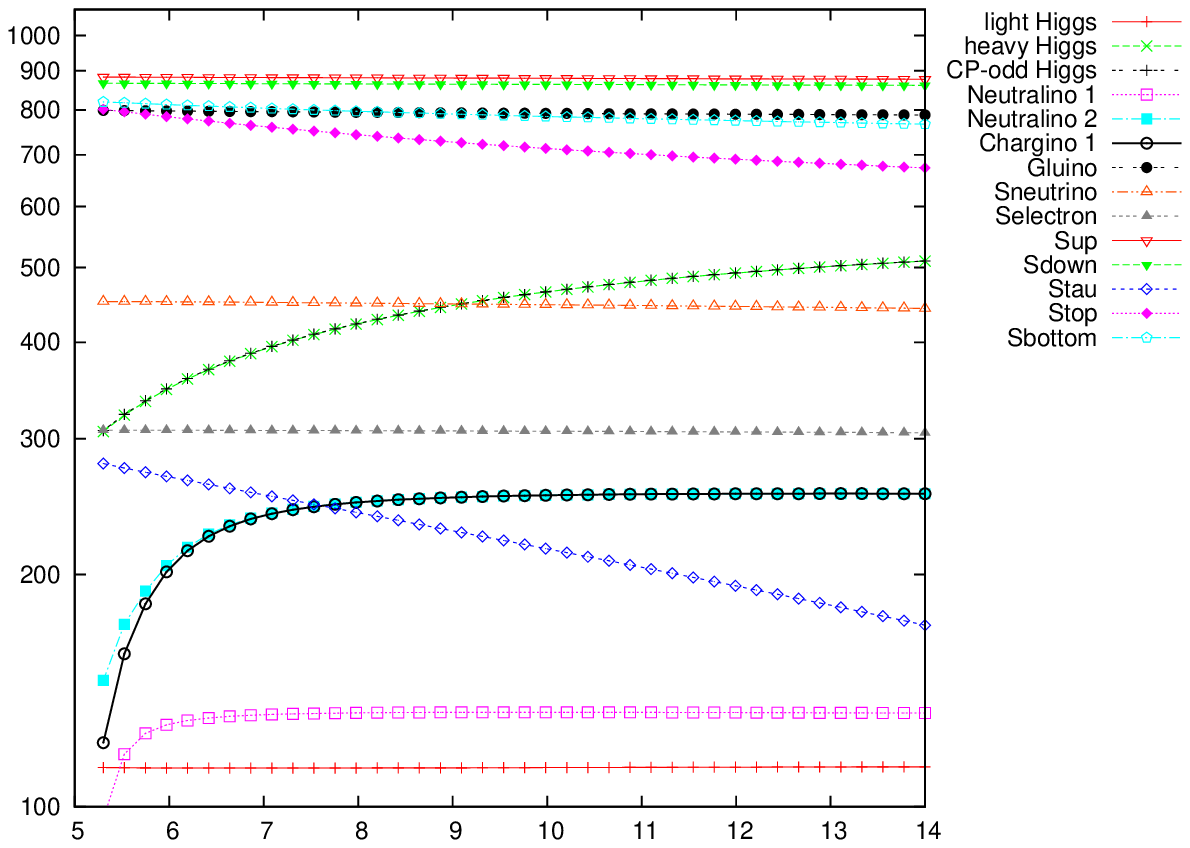}
\rText(-365,150)[c][l]{\scalebox{1}[1]{Mass in GeV}}
\Text(-200,-10)[c]{\scalebox{1}[1]{$\log_{10} \left(M_{\rm mess}/{\rm GeV}\right)$}}
\end{picture}
}
\vspace*{0.5cm}
\caption{The particle spectrum resulting from $ \Lambda_{S,r}(M_{\rm mess}) = \Lambda_{G,r} = 10^{5}\text{GeV}$ held constant as $M_{\rm mess}$ is
varied (upper panel). In the lower panel we show the spectrum for $ \Lambda_{S,r}(M_{\rm mess})$ varied as in Fig.~\ref{unification}, such that the $I_{M_{r}}$ RGIs remain constant. $\tan\beta = 45$}
\label{fig:spectraTanb45}
\end{figure}

\begin{figure}[t]
\centering
\begin{picture}(330,260)
\includegraphics[width=0.65\linewidth]{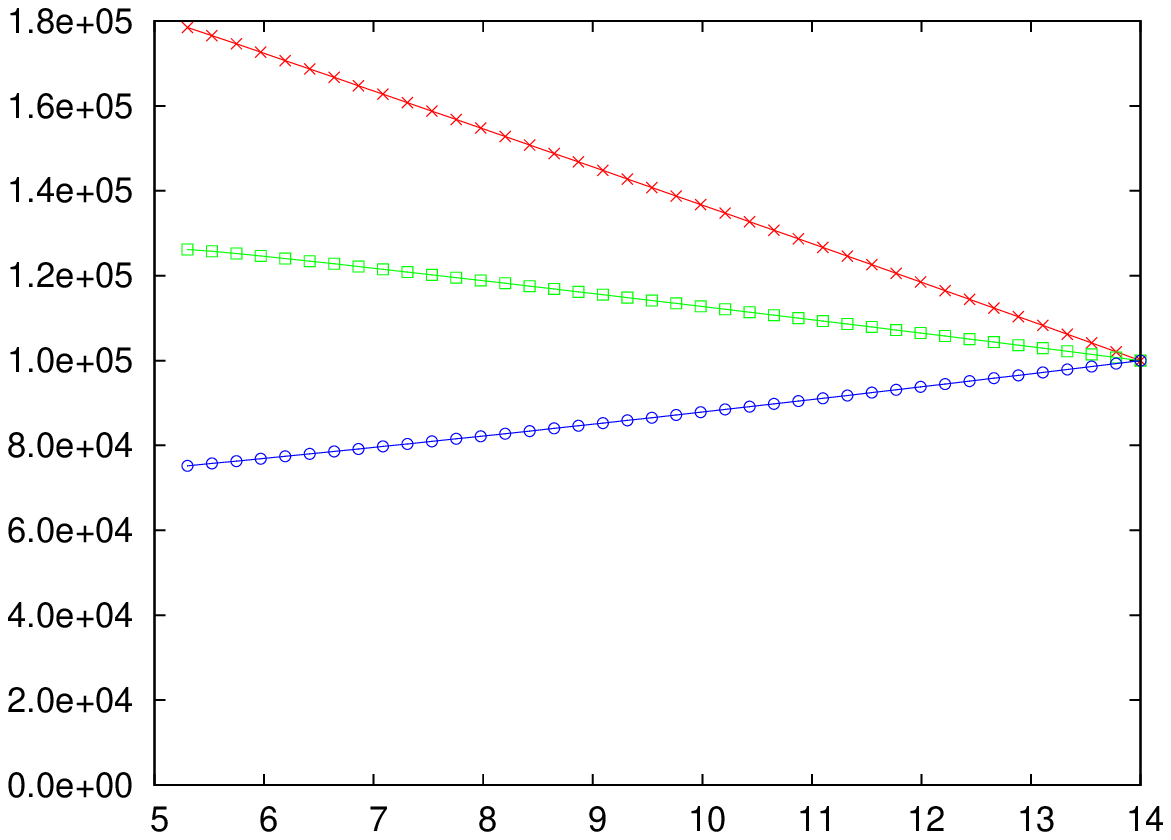}
\rText(-300,110)[c][l]{\scalebox{1}[1]{$\Lambda_{S,r}$ in GeV}}
\Text(-130,-10)[c]{\scalebox{1}[1]{$\log_{10} \left(M_{\rm mess}/{\rm GeV}\right)$}}
\end{picture}
\vspace*{0.5cm}
\caption{Values of $\Lambda_{S,r}$ as functions of the messenger scale, keeping the RGIs fixed.
We chose the point in parameter space with all six $\Lambda$s equal, $\Lambda_{S,r}=\Lambda_{G_r}=10^5\, {\rm GeV}$, at the
messenger scale $M_{\rm mess}=10^{14}\, {\rm GeV}$. For lower values of $M_{\rm mess}$ different values of $\Lambda_{S,r}$ are required
to keep the same values of the $I_{M_r}$ RGIs in Table~\ref{rginvariants}. These $\Lambda_{S,r}(M_{\rm mess})$ are plotted here: from
top to bottom, $\Lambda_{S,1}$ (in red), $\Lambda_{S,2}$ (in green), and $\Lambda_{S,3}$ (in blue). }
\label{unification}
\end{figure}

In Fig.~\ref{fig:spectraTanb45} we first compare two GGM models while varying the messenger scale. In the upper panel we have kept
all $\Lambda$s fixed while varying $M_{\rm mess}$, and as one can see the spectrum changes significantly.
In the lower panel instead of fixing the $\Lambda$s we keep the values of the RGIs constant as $M_{\rm mess}$ is varied
(the values of $\Lambda_{S,r}$ as functions of $M_{\rm mess}$ required to achieve this are shown in Fig.~\ref{unification}).
Here, one can clearly see that the first and second generation sfermion masses show basically no variation.
From this perspective the situation is considerably improved from the upper to the lower panel of Fig.~\ref{fig:spectraTanb45}.
We can, however, also see that the third generation sfermion masses still change with $M_{\rm mess}$.

\begin{figure}[t]
\centering
\begin{picture}(350,260)
\includegraphics[width=0.75\linewidth]{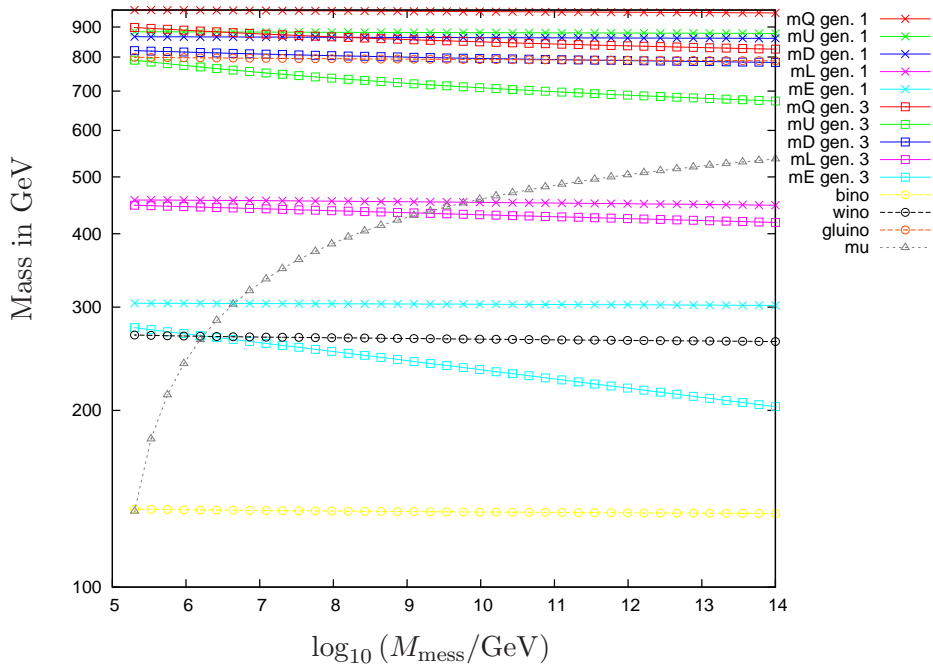}
\rText(-350,145)[c][l]{\scalebox{1}[1]{Mass in GeV}}
\Text(-195,-10)[c]{\scalebox{1}[1]{$\log_{10} \left(M_{\rm mess}/{\rm GeV}\right)$}}
\end{picture}
\vspace*{0.5cm}
\caption{Soft mass parameters for the models chosen as in the lower panel of Fig.~\ref{fig:spectraTanb45}.
Solid lines with crosses (boxes) correspond to first (third) generation sfermions. Dashed lines show the bino, wino and gluino masses and the 
$\mu$-parameter.}
\label{fig:gauginosoft}
\end{figure}

The gluino mass also stays constant as dictated by the corresponding RGI.
In contrast the neutralino and chargino masses in the lower panel of Fig.~\ref{fig:spectraTanb45} seem to depend quite strongly on
the messenger scale.  However, these masses do not only depend on the soft gaugino mass parameters $M_{1}$ and $M_{2}$ but also on the details of
electroweak symmetry breaking.
Indeed the neutralino mass matrix can be written as,
\begin{equation}
\label{neutmix}
M_{\tilde{N}}=\left ( \begin{array}{cccc}
    M_{1} & 0 & -c_{\beta}s_{W}m_{Z} & s_{\beta}s_{W}m_{Z} \\
    0 & M_{2} & c_{\beta}c_{W}m_{Z} & -s_{\beta}c_{W}m_{Z} \\
    -c_{\beta}s_{W}m_{Z} & c_{\beta}c_{W}m_{Z} & 0 & -\mu \\
    s_{\beta}s_{W}m_{Z} & -s_{\beta}c_{W}m_{Z} & -\mu & 0 \\
  \end{array}\right),
\end{equation}
where we have used the abbreviations $s_{\beta}=\sin\beta$, $c_{\beta}=\cos\beta$, $s_{W}=\sin\theta_{W}$ and $c_{W}=\cos\theta_{W}$
with the Weinberg angle $\theta_{W}$.
We can now use this to determine the soft masses $M_{1}$ and $M_{2}$ as well as $\mu$ from the
neutralino masses and mixings (one could use a similar procedure with the chargino masses).
The results for the same model as in the lower panel of Fig.~\ref{fig:spectraTanb45} are shown in Fig.~\ref{fig:gauginosoft}. As we can see the gaugino mass
parameters  (as well as those for the first generation sfermions) are constant to good accuracy, as expected from the RGIs. The neutralino mass dependence arises entirely from
the variation of $\mu$. 

From Fig.~\ref{fig:gauginosoft}, in which $\tan\beta=45$, we can see that all third generation sfermion soft mass parameters depend on $M_{\rm mess}$. For low values of $\tan\beta$ the third generation $m^{2}_{\tilde{L},3}$, $m^{2}_{\tilde{b}}$ and $m^{2}_{\tilde{\tau}}$ soft parameters 
equal those of the first generation $m^{2}_{\tilde{L},1}$, $m^{2}_{\tilde{d}}$ and $m^{2}_{\tilde{e}}$ and the former become independent of $M_{\rm mess}$, too. 
This is in line with the fact that in low $\tan\beta$ models, $y_b$ and $y_{\tau}$ are decreased in line with a correspondingly 
greater $\langle H_d \rangle$.  The two third generation soft parameters $m^{2}_{\tilde{Q},3}$ and $m^{2}_{\tilde{t}}$ continue to depend on $M_{\rm mess}$ since the top Yukawa remains large.

The six non-vanishing RGIs (whose low scale values are equal to their values at the high, messenger scale) extracted from the first generation sfermion masses as well as the neutralino/chargino masses and mixings uniquely and directly determine the high scale $\Lambda$s for a
given messenger scale. Looking only at these quantities we would think that we indeed have a set of equivalent models.
However, including also third generation sfermions as well as the $\mu$ parameter makes it clear that these models which
differ only in the value of the messenger scale give indeed quite different physical spectra. This shows that $M_{\rm mess}$
is indeed a true parameter.

As we have seen $M_{\rm mess}$ most directly influences the third generation, which has special significance as it gives the lightest sfermions.
Furthermore, neutralino masses are also significantly affected by the value of $M_{\rm mess}$. This is because of the sizable variation 
of $\mu$.
As a supersymmetric parameter $\mu$ does not really evolve with the RG flow.
However, as we keep the parameters of electroweak symmetry breaking (in particular $m_{Z}$ and $\tan\beta$) fixed, $\mu$ is determined a posteriori from these as well as the soft Higgs mass parameters $m^{2}_{H_{u}}$ and $m^{2}_{H_{d}}$. It is the non-trivial dependence of the latter on the messenger scale that leads to the dependence of~$\mu$ on $M_{\rm mess}$ shown in Fig.~\ref{fig:gauginosoft}.

We can also see that the messenger scale is a true parameter directly from the 1-loop RG equations. 
A necessary, but not sufficient condition for the equivalence of two models is that their RGIs agree (after all RGIs are made from observables).
Now for GGM the six RGIs, $I_{B_{r}}$ and $I_{M_{r}}$,
uniquely determine all $\Lambda$s for a given messenger scale. 
If the messenger scale was not a true parameter of GGM, two points in the parameter space with the same six RGIs but different values of the
messenger scale $M_{\rm mess}$ and $M^{\prime}_{\rm mess}$ would have to be physically equivalent. In particular all of their observables, not just the RGIs, would have to agree.
The evolution of one such parameter is particularly easy to follow.
This is the third generation analog, $I_{M_{r},3}$ of the $I_{M_{r}}$ (defined in Sect.~\ref{rgi} above Eq.~\eqref{RGvariants}).
At each messenger scale the expression for $I_{M_{r},3}$ in terms of the $\Lambda_{S,r}$ is the same as $I_{M_{r}}$ and consequently
$I_{M_{r},3}(M_{\rm mess})=I_{M_{r},3}(M^{\prime}_{\rm mess})$. 
However, as we can see from Eq.~\eqref{RGvariants} the $I_{M_{r},3}$ change
during the RG evolution. Evolving between the two models (from $M_{\rm mess}$ to $M^{\prime}_{\rm mess}$) we find that the $I_{M_{r},3}$
do not agree between the two scales due to the non-negligible third generation Yukawa couplings.  
This demonstrates that $M_{\rm mess}$ does not just give a re-parameterisation of the parameter space but is indeed a true additional
input.

We note that this argument applies not just to GGM but to any flavour-blind model of supersymmetry breaking: at the high scale 
$I_{M_r} = I_{M_r,3}$, but the latter runs whereas the former does not, so the third generation sfermions at low scale will always depend on what the high-scale was.

\section{Unification of sfermion mass parameters}\label{sec:unification}
Let us now turn to a phenomenological application of the RGIs in the context of GGM.

\subsection{Unification strategy}
There is a relation between non-vanishing RGIs and the input soft parameters at the high scale. This is given in the third column
of Tab.~\ref{rginvariants}. In particular there are eight soft parameters in GGM: three $\Lambda_{S, r}$, three $\Lambda_{G,r}$
and two $\delta_{u,d}$.
The latter are given by,
\begin{equation}
\label{useless}
\delta_{u}=-\frac{1}{2}\left(D_{Z}+\frac{13}{5}D_{Y_{13H}}\right)\, ,\quad\quad \delta_{d}=-\frac{1}{2}D_{Z}\, ,
\end{equation}
as can be easily seen from Tab.~\ref{rginvariants}.
The $\Lambda$s can be similarly determined in a straightforward manner,
\begin{equation}
\label{lambdas}
\Lambda^{2}_{S,r}   =  \frac{(16 \pi^{2})^2}{\kappa_{r} k_{r}^2} \left( \frac{I_{M_{r}}}{ g_{r}^{4} } - I_{B_{r}}^{2} \right)
\end{equation}
where we defined $\kappa_{r} = (33/5, 1, -3)$, and $k_{r} = (5/3, 1, 1)$ as before.
Similarly,
\begin{equation}
\label{lambdag}
\Lambda_{G,r}=I_{B_{r}}\frac{16\pi^2}{k_{r}}.
\end{equation}

Without loss of generality, but for simplicity of presentation we will mostly focus on models with $\delta_{u}=\delta_{d}=0$.
In these models $D_{Z}=D_{Y_{13H}}=I_{Y\alpha}=0$.
The non-vanishing set of RGIs in these GGM models collapses to the six quantities: $I_{B_{r}}$, $I_{M_{r}}$.

At the high scale, these six parameters uniquely determine the six $\Lambda$s. Assuming constancy of the six RGIs, 
Eqs.~\eqref{lambdas} and \eqref{lambdag} determine values of $\Lambda_{S}(M)$ and $\Lambda_{G}(M)$ at any scale $M$. 
In this sense we have a concept of running $\Lambda$s. An example of this can be seen in Fig.~\ref{unification}. The variation of RGIs due to two-loop running, and the resulting effect
on the running $\Lambda$s is discussed at the end of the section.

As discussed in the previous section, had there been no effect of the third generation on the model observables (such as the mass spectrum)
any model with the same values of the six RGIs would be equivalent. The running $\Lambda$s obtained from these RGIs at different
values of the messenger scale would then define a set of equivalent models. Different values of the messenger scale would
then just lead to a re-parameterisation of the GGM parameter space\footnote{A set of equations to give a re-parameterisation in model space resulting in a fixed spectrum for the first and second generation as well as gauginos has been given in~\cite{Carena:2010wv}.}.
As we have seen above, the third generation lifts this model degeneracy.

Now we are ready to address the notion of unification in terms of $\Lambda$s.
As we will see shortly this originates from Grand Unification but goes beyond the usual unification of gauge couplings and gives
quite powerful complementary information.

First let us assume that the theory grand unifies. We further assume that there is a single messenger in a complete GUT representation
with a mass $M_{\rm mess}$. In other words we are assuming a complete GUT multiplet with no doublet-triplet-like splitting.
In this case one finds that Eqs.~\eqref{gauginosoft}, \eqref{scalarsoft} hold at the scale $M_{\rm mess}$ and there is a single
\begin{equation}
\label{unified1}
\Lambda_{S}=\Lambda_{S,1}=\Lambda_{S,2}=\Lambda_{S,3}
\end{equation}
 and a single
\begin{equation}
\label{unified2}
\Lambda_{G}=\Lambda_{G,1}=\Lambda_{G,2}=\Lambda_{G,3}.
\end{equation}
In other words, there is no splitting in the $\Lambda$s between the different gauge factors.

Starting from unified $\Lambda_{S}$, $\Lambda_{G}$ as in Eqs.~\eqref{unified1}, \eqref{unified2} and using Eqs.~\eqref{lambdas} and \eqref{lambdag} to define running $\Lambda_{S,r}(M)$ and $\Lambda_{G,r}(M)$, we find that at scales below
$M_{\rm mess}$ the $\Lambda_{S,r}(M)$ take different values for the different gauge group factors due to the presence of the running couplings
in Eq.~\eqref{lambdas}.
The $\Lambda_{G,r}(M)$, however, remain unsplit (indeed they do not run).
The splitting of the $\Lambda_{S,r}$ is shown in Fig.~\ref{unification}
where the horizontal scale can be equivalently thought of as an RG scale $M$.

Now let us consider the picture from the low scale. From the observable masses one can construct the RGIs. Again, assuming that
these RGIs are constant to a good approximation, we
can use Eqs.~\eqref{lambdas} and \eqref{lambdag} to extract the six $\Lambda$ parameters at any scale.
A priori we do not know whether the theory unifies (in the $\Lambda$ sense described above).
However if it turns out that all the three $\Lambda_{G,r}$ match, this is the first indication that unification can occur.
The next step is to examine the running of the three $\Lambda_{S,r}$. True unification corresponds to the behavior shown in
Fig.~\ref{unification} where the three lines meet at one point.
If this is the case, it gives a very strong indication for a unified structure
and at the same time also predicts the fundamental value of the messenger scale.

The unified model has a reduced set of fundamental parameters. There are essentially three of them:
$\Lambda_{S}$, $\Lambda_{G}$ and $M_{\rm mess}$ (exactly as in the setup of~\cite{ADJKpGGM,ADJK7}) plus information about
either $\tan\beta$ or $B_{\mu}$.
From these fundamental parameters at the now known messenger scale one can compute the entire mass spectrum.
In particular the third generation data, which has not been used in the extraction of $\Lambda_{S}$, $\Lambda_{G}$ and $M_{\rm mess}$ can now be used
as a non-trivial cross-check of the model with the data.

On the other hand if no signs of the unification described above are observed, this is an indication of either
a non-gauge-mediated structure or the presence of doublet-triplet splitting effects in the messenger sector.
In this case one can still test whether we are dealing with a GGM model, a more general flavour blind model or neither. 
For GGM all three RGIs $D_{B_{13}}$,
$D_{L_{13}}$ and $D_{\chi_{1}}$ have to vanish. However, flavour blindness on its own does not require the vanishing of $D_{\chi_{1}}$ and
therefore constitutes a test of GGM~\cite{Carena:2010gr}.

The whole discussion above can be easily generalised to the case of non-vanishing $\delta_{u}$ and $\delta_{d}$ defined 
in Eqs.~\eqref{higgssoft}. These quantities can be extracted from the RGIs $D_{Z}$, $D_{Y_{13H}}$ and $I_{Y\alpha}$.
The non-vanishing of any of these implies that additional contributions to the Higgs masses are present. The non-vanishing
of $D_{Y_{13H}}$ and $I_{Y\alpha}$ points to non-universal Higgs masses.

\subsection{Effects of experimental and theoretical errors}
The use of the RGIs to determine the running $\Lambda_S$ parameters (and the possible unification thereof) relies upon the accuracy with which
they can be determined experimentally, and also on the extent to which they remain constant during RG evolution.
In terms of the soft mass parameters\footnote{Above we have expressed the running $\Lambda_{S,r}$ in terms of the RGIs, however
the error in $I_{B_{r}}$ is correlated with that in $I_{M_{r}}$.} determined at the low scale 
(parameters without argument are evaluated at the low scale: 
$m_{\tilde{f}}=m_{\tilde{f}}|_{\rm low \,scale}$, $M_{r}=M_{r}|_{\rm low \,scale}$ and $g_{r}=g_{r}|_{\rm low \,scale}$),
the $\Lambda_{S}$ parameters at the scale $M$ are given by,
\begin{eqnarray}
\label{lambdadetermination}
\frac{\Lambda^{2}_{S,1}(M)}{(16 \pi^{2})^2}  & =&  \frac{3}{55}  \left[ M_1^2 \left( \frac{1}{g_{1}^{4}(M)} - \frac{1}{g_{1}^{4}} \right) - \frac{1}{g_{1}^{4}(\mu)} \frac{33}{8}(m_{\tilde{d}_1}^2-m_{\tilde{u}_1}^2-m_{\tilde{e}_1}^2) \right] \\\nonumber
\frac{\Lambda^{2}_{S,2} (M)}{(16 \pi^{2})^2} & = &  M_2^2 \left( \frac{1}{g_{2}^{4}(M)} - \frac{1}{g_{2}^{4}} \right) + \frac{1}{g_{2}^{4}(\mu)} \frac{1}{24}\left(9(m_{\tilde{d}_1}^2-m_{\tilde{u}_1}^2)+16m_{\tilde{L}_1}^2-m_{\tilde{e}_1}^2\right)  \\\nonumber
\frac{\Lambda^{2}_{S,3} (M)}{(16 \pi^{2})^2} & = &-\frac{1}{3}  \left[ M_3^2 \left( \frac{1}{g_{3}^{4}(M)} - \frac{1}{g_{3}^{4}} \right) - \frac{1}{g_{3}^{4}(\mu)} \frac{3}{16}(5m_{\tilde{d}_1}^2+m_{\tilde{u}_1}^2-m_{\tilde{e}_1}^2) \right].
\end{eqnarray}
The experimental errors arise from the imprecise determination of the soft mass parameters at the 
low scale, $\Delta m^{2}_{\tilde{f}}|_{\rm low\, scale}$ and $\Delta M_{r}|_{\rm low\, scale}$.
We note that the squark and slepton
masses appearing 
above are those of the first (and second) generation on which electroweak symmetry breaking
has negligible effect, and consequently one can use physical masses directly.
One has to be more careful, however, with the gaugino
masses, $M_r$, 
which have the meaning of soft mass terms for gauginos, distinct from
the physical chargino and neutralino masses (gluinos on the other hand do not mix with any other particle). Determination of $M_1$ and $M_2$ thus requires measurement of the
relevant mixing angles that diagonalise Eq.~\eqref{neutmix}. In addition, as discussed in \cite{Carena:2010gr}, accurate measurement of
$I_{M_1}$ and consequently $\Lambda_{S,1}$ could prove difficult in models with relatively heavy squarks due to the appearance of $m_{\tilde{d}_{1}}^2-m_{\tilde{u}_{1}}^2$ multiplied by a large coefficient.

\begin{figure}
\centering
\subfigure{
\begin{picture}(200,200)
\includegraphics[width=0.45\linewidth]{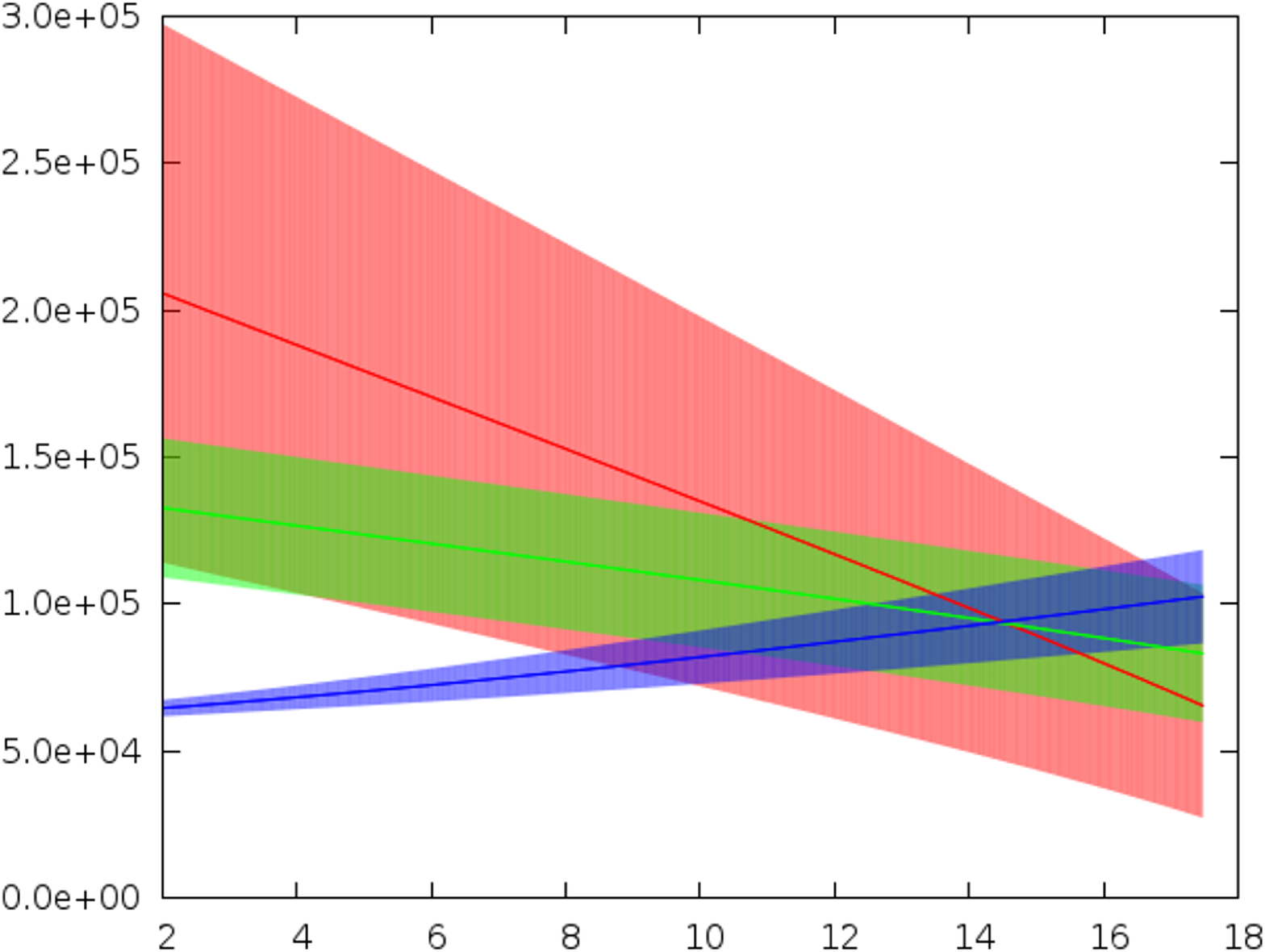}
\rText(-210,80)[c][l]{\scalebox{0.7}[0.7]{$\Lambda_{S,r}$ in GeV}}
\Text(-90,-5)[c]{\scalebox{0.7}[0.7]{$\log_{10} \left(M_{\rm mess}/{\rm GeV}\right)$}}
\end{picture}
}
\hspace*{0.7cm}
\subfigure{
\begin{picture}(200,200)
\includegraphics[width=0.45\linewidth]{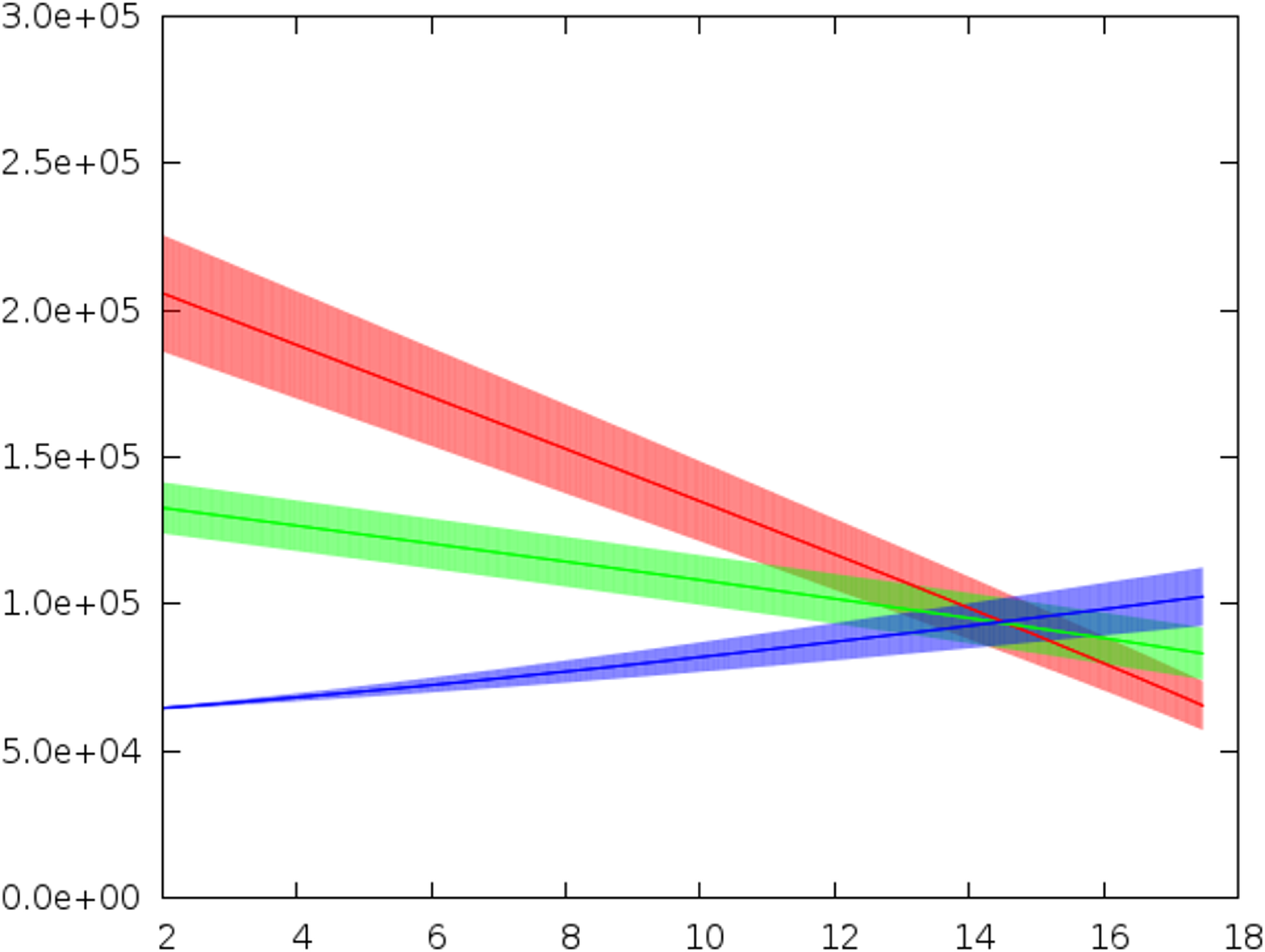}
\rText(-210,80)[c][l]{\scalebox{0.7}[0.7]{$\Lambda_{S,r}$ in GeV}}
\Text(-90,-5)[c]{\scalebox{0.7}[0.7]{$\log_{10} \left(M_{\rm mess}/{\rm GeV}\right)$}}
\end{picture}
}
\vspace*{0.4cm}
\caption{Reconstructed values of $\Lambda_{S,r}$ with experimental errors shown as shaded bands. The left panel corresponds to 
a $5\%$ error in the determination of all soft masses. The right panel corresponds to a $5\%$ uncertainty in all soft masses apart from
$m_{\tilde{u}_{1}}$ and $m_{\tilde{d}_{1}}$ which are assumed to be determined to $1\%$. The parameters of the model are chosen as in Fig.~\ref{unification}.}
\label{fig:bands}
\end{figure}

The experimental uncertainties $\Delta m^{2}_{\tilde{f}}|_{\rm low\, scale}$ result in uncertainties for the $\Lambda_{S}$ parameters which are shown as colored bands in Fig.~\ref{fig:bands}. We see that the determination of the messenger scale and the clarity of the unification
pattern crucially depends on an accurate measurement of $m_{\tilde{u}_{1}}^2$ and $m_{\tilde{d}_{1}}^2$ or more precisely
$m_{\tilde{d}_{1}}^2-m_{\tilde{u}_{1}}^2$.

\begin{figure}[!hp]
\centering
\subfigure{
\scalebox{0.8}[0.8]{\begin{picture}(280,190)
\includegraphics[width=0.6\linewidth]{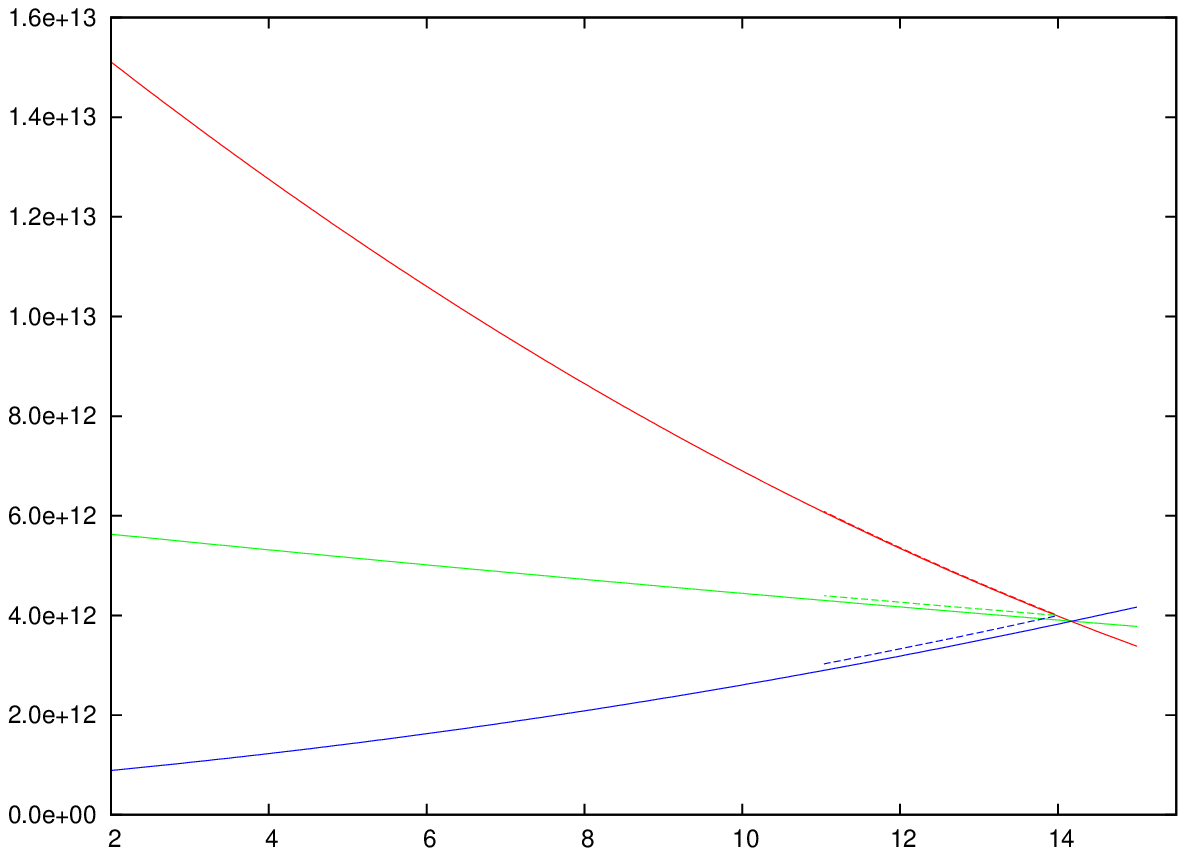}
\CArc(-36,54)(2,0,360)
\SetColor{Red}
\CArc(-32,52.5)(2,0,360)
\rText(-280,98)[c][l]{\scalebox{0.8}[0.8]{$\Lambda_{S,r}$ in GeV}}
\Text(-115,-10)[c]{\scalebox{0.8}[0.8]{$\log_{10} \left(M_{\rm mess}/{\rm GeV}\right)$}}
\end{picture}
}
}
\\[0.5cm]
\subfigure{
\scalebox{0.8}[0.8]{
\begin{picture}(280,190)
\includegraphics[width=0.6\linewidth]{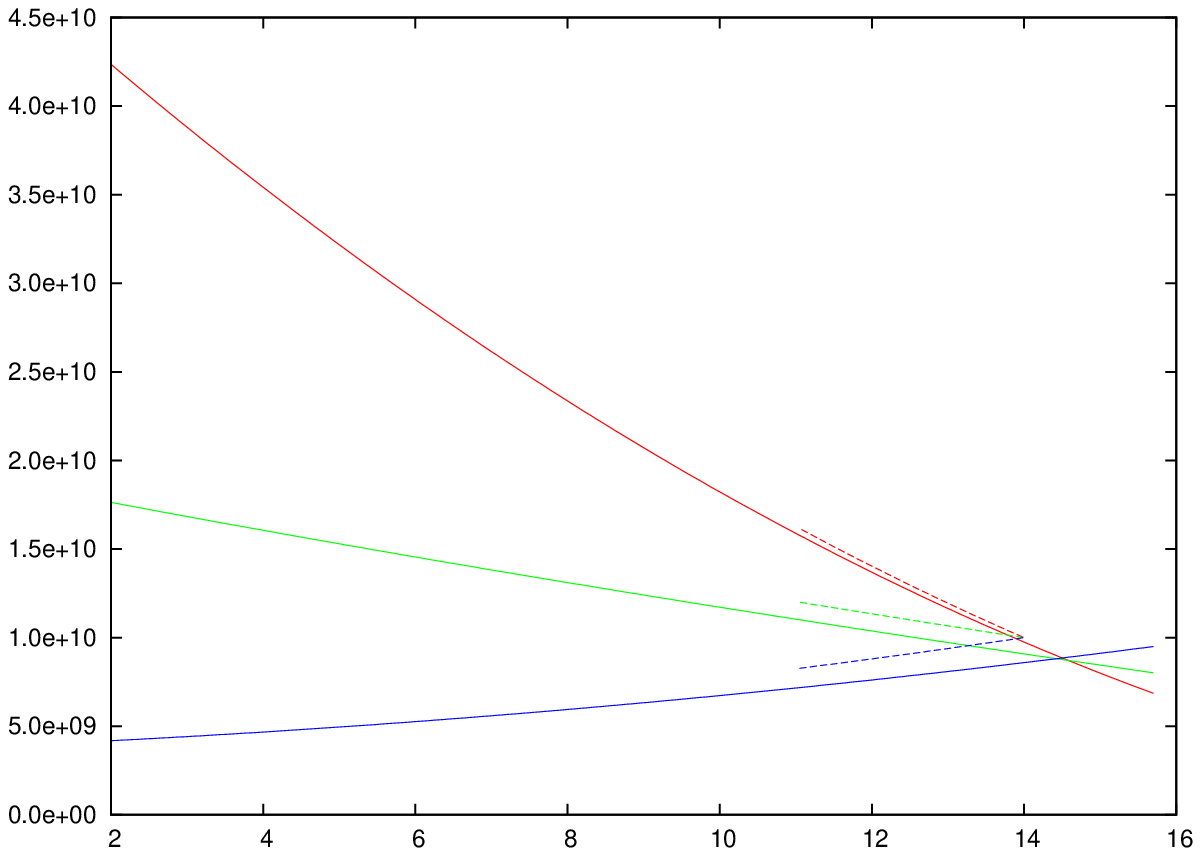}
\CArc(-43,49.5)(2,0,360)
\SetColor{Red}
\CArc(-34,44.5)(2,0,360)
\rText(-280,98)[c][l]{\scalebox{0.8}[0.8]{$\Lambda_{S,r}$ in GeV}}
\Text(-115,-10)[c]{\scalebox{0.8}[0.8]{$\log_{10} \left(M_{\rm mess}/{\rm GeV}\right)$}}
\end{picture}
}
}\\[0.5cm]
\subfigure{
\scalebox{0.8}[0.8]{
\begin{picture}(280,190)(-7,0)
\includegraphics[width=0.587\linewidth]{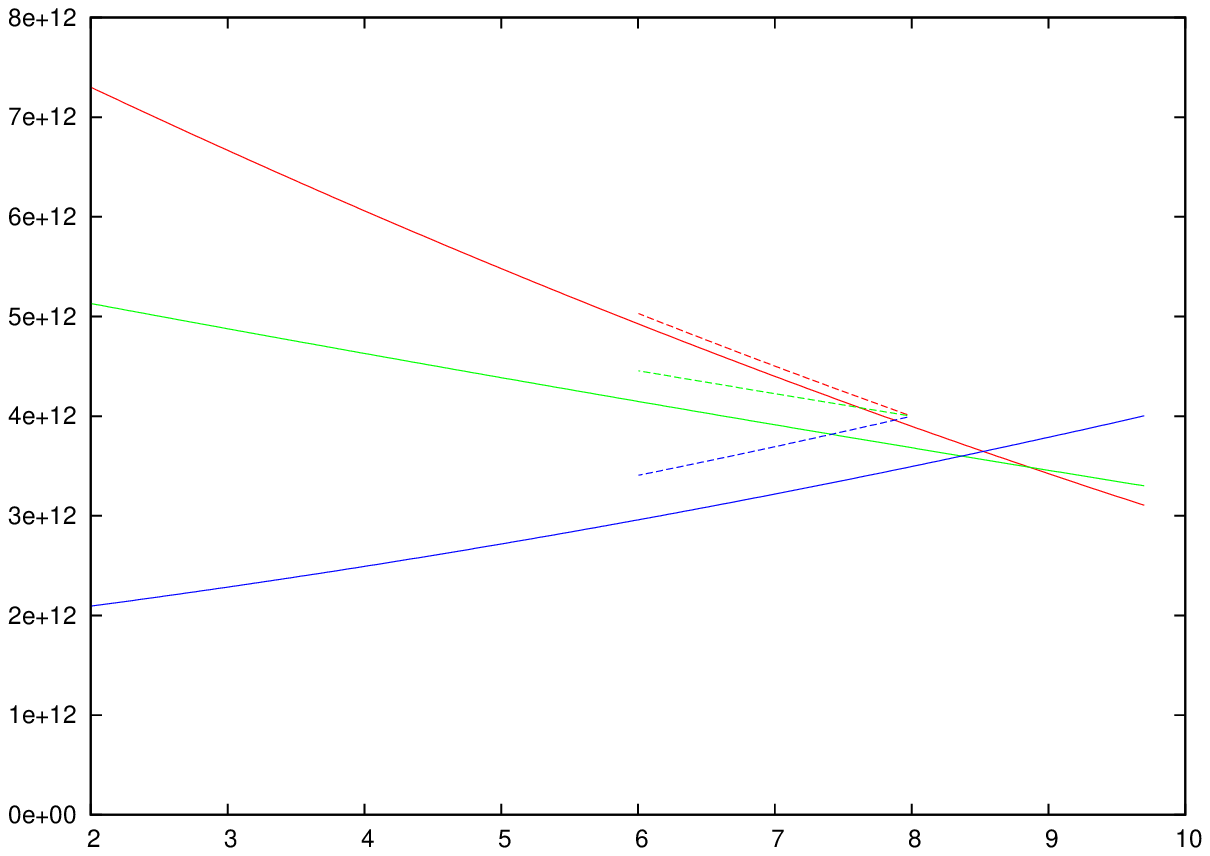}
\CArc(-68.5,95.5)(2,0,360)
\SetColor{Red}
\Oval(-49,86)(4,10)(0)
\rText(-280,98)[c][l]{\scalebox{0.8}[0.8]{$\Lambda_{S,r}$ in GeV}}
\Text(-115,-10)[c]{\scalebox{0.8}[0.8]{$\log_{10} \left(M_{\rm mess}/{\rm GeV}\right)$}}
\end{picture}
}
}\\[0.5cm]
\caption{The running $\Lambda_{S,r}$ reconstructed from the low-scale values of the RGIs in three GGM models
with high-scale unification. The top panel has $\Lambda_G = 5 \times 10^{5}\, {\rm GeV}$
and $\Lambda_S= 2 \times 10^{6}\, {\rm GeV}$ at $M_{\rm mess} = 10^{14}\, {\rm GeV}$.
The model in the middle panel is the same as the model used before for Fig.~\ref{unification}.
The lower panel corresponds to $\Lambda_G = \Lambda_S= 2 \times 10^{6}\, {\rm Gev}$ at $M_{\rm mess} = 10^{8}\, {\rm GeV}$.
In all three cases the running $\Lambda_{S,r}$ are insensitive to the value of $\tan\beta$.
The black dot indicates the true unification point. The region where solid lines (nearly) intersect gives the reconstructed unification.}
\label{fig:theor}
\end{figure}

Let us now investigate how the theoretical error can affect the unification picture as seen from the low scale.
For concreteness, let us consider models where the high-scale unification of the $\Lambda_S$ parameters does occur.
Figure~\ref{fig:theor} shows three representative models of this type. The dotted lines show the three $\Lambda_S$ parameters
at the true messenger scale $M_{\rm mess}$ and just below, showing their unification at $M_{\rm mess}$.
The model is then evolved numerically (to two-loop precision) to the low scale where the mass spectrum is determined.
From this mass spectrum we compute the RGIs and thence the $\Lambda_S$  parameters at the low scale. The running $\Lambda_S$ parameters are then determined at higher scales through Eqs.~\eqref{lambdas}, assuming the RGIs do not deviate from their low-scale values.
These reconstructed running $\Lambda_S$ parameters are plotted as solid lines.
In all of the examples considered this simple procedure maintains the unification pattern of the $\Lambda_{S}$ parameters and also gives a good estimate for the value of the underlying messenger scale. 

As we can see by comparing Figs.~\ref{fig:bands} and \ref{fig:theor} the projected experimental uncertainties dominate.
If at some later point in time the experimental precision of measuring sparticle masses can be significantly improved 
one would also be able to remove the theoretical uncertainty discussed here by going beyond the one-loop RGI treatment.
More precisely, one could use the GGM parameter values estimated from the RGI treatment of unification and numerically (using e.g.
{\texttt SoftSUSY}) explore 
a region around this estimate for the unification point.

\section{Conclusions}
Assuming SUSY is discovered, one of the main questions for SUSY phenomenology 
will be how to ultimately discern the mediation mechanism of SUSY breaking and the structure of the underlying theory at the high scale.
In this paper we have attempted to answer this question in the context of general gauge mediation (GGM) by looking for imprints
of the theory at the high scale on the observables at the electroweak scale. 
In particular we have proposed a procedure of searching for unification of the scalar mass parameters $\Lambda_{S; 1,2,3}$ appearing in the context of general gauge mediation. In this case one can draw powerful conclusions about unification itself as well
as the value of the underlying messenger scale at which unification occurs. This goes beyond the usual unification of gauge couplings
at $M_{\rm GUT}$ and gives information on the structure of the messengers -- they should appear in complete and unsplit multiplets
multiplets for the unification to hold -- as well as their masses.
In this situation the underlying parameter space is restricted to just a unique $\Lambda_{S}(M_{\rm mess})$ scalar mass parameter,
a unique $\Lambda_{G}(M_{\rm mess})$ gaugino mass scale, the messenger mass $M_{\rm mess}$ itself (as well as $\mu$, $\tan\beta$ and possibly $\delta_{u}$ and $\delta_{d}$ Higgs mass contributions -- all of which can in principle be determined from the observables at the low scale). Establishing this unification pattern (and determining the additional parameters) requires a sufficiently precise determination of sparticle masses. We have investigated the effects of experimental and theoretical uncertainties on the conclusions one can draw. Theoretical errors can easily be brought under control. Requirements on the experimental accuracy are fairly strong.

We have also shown that the messenger mass is an active and phenomenologically relevant parameter in GGM whether or not the unification of the $\Lambda_{S}$ mass scales occurs.

\subsection*{Acknowledgements}
We are grateful to Matt Dolan for useful discussions
and comments.


\providecommand{\href}[2]{#2}\begingroup\raggedright
\endgroup

\end{document}